\documentclass[twocolumn]{aastex701}

\begin{document}

\title{The radial velocity curve for \ion{He}{2} emission cannot be used for component mass determination in SS433}

\author[orcid=0000-0002-6755-2120,sname='Dodin']{A.V. Dodin}
\affiliation{Sternberg Astronomical Institute, Lomonosov Moscow State University, Universitetskij prospekt 13, 119234 Moscow, Russia}
\email[show]{samsebedodin@gmail.com, kpostnov@gmail.com, \\ cherepashchuk@gmail.com}  

\author[orcid=0000-0002-1705-617X, sname='Postnov']{K.A. Postnov} 
\affiliation{Sternberg Astronomical Institute, Lomonosov Moscow State University, Universitetskij prospekt 13, 119234 Moscow, Russia}
\email[noshow]{kpostnov@gmail.com}

\author[orcid=0000-0001-5595-2285, sname='Cherepashchuk']{A.M. Cherepashchuk}
\affiliation{Sternberg Astronomical Institute, Lomonosov Moscow State University, Universitetskij prospekt 13, 119234 Moscow, Russia}
\email[noshow]{cherepashchuk@gmail.com}

\begin{abstract}
More than 150 measurements of the \ion{He}{2} 4686\,{\AA} emission line in spectra of  SS433 were obtained during 388 nights in $2020-2025$ with the Transient Double-beam Spectrograph on the 2.5 m telescope of Caucasian Mountain Observatory of Sternberg Astronomical Institute. We found that the \ion{He}{2} emission line formation region is not  eclipsed and is significantly larger than both the donor star and the photosphere of the supercritical accretion disk. The \ion{He}{2} radial velocity curve was found to be independent of the precessional phase and inconsistent with the photometric curve. These findings suggest that the \ion{He}{2} line does not reflect the orbital motion of the compact object.  Therefore, spectroscopic estimates of the masses of the components in SS433 based on the \ion{He}{2} emission line can be unrealistic.
\end{abstract}


\section{Introduction}

The Galactic microquasar SS433 is a massive eclipsing X-ray binary system at an advanced evolutionary stage with a supercritical accretion disk around a black hole and relativistic jets \citep{1984ARA&A..22..507M,2004ASPRv..12....1F,2020NewAR..8901542C,2025arXiv250601106C}. The object displays three variabilities: orbital ($P_{ \rm orb} \approx13^d.1$), precessional ($P_{ \rm prec} \approx162^d.3$), and nutational ($P_{ \rm nut}=6^d.29$) ones. Spectroscopic studies of SS433 as a close binary system have been carried out by many authors. From  Doppler shifts of narrow components of stationary hydrogen emission lines, a period of $13^d.1$ was detected, suggesting that SS433 is a close binary system \citep{1980ApJ...235L.131C}. In the paper by \cite{1981MNRAS.194..761C}, optical eclipses were discovered in the SS433 system and it was shown that radial velocity curves constructed from stationary hydrogen emission lines \citep{1980ApJ...235L.131C} do not reflect the orbital motion of the components since their shape is inconsistent with the positions of the primary and secondary minima on the light curve.

The discovery of double-peaked profiles of stationary  hydrogen and neutral helium emission lines suggested the existence of a circumbinary shell around SS433 rotating with a velocity of about 200 km s$^{-1}$ \citep{1988AJ.....96..242F, 2011A&A...531A.107B}. The presence of a circumbinary disk with this rotation velocity was confirmed by interferometric measurements in the Br$\gamma$ line at a distance of $\sim5$ au from the central source \citep{2019A&A...623A..47W}, which implies a very high total mass of SS433 ($\gtrsim400$M$_{\odot}$) assuming the Keplerian rotation. Other observations presented in the same paper, but performed at a different time, show that the line profiles are dominated by a bipolar outflow. Despite such a complex and variable formation region of hydrogen lines, their average radial velocity reflects the orbital periodicity of the binary system, but does not provide information about the radial velocities of the components.
Actually, the fact that the maximum of the radial velocity occurs in the middle of the eclipse reflects the flux-averaged motion of different emitting regions, including the powerful gas flow feeding the accretion disk or matter escaping through the outer Lagrangian point \citep{2023MNRAS.519.1409L}. 

Starting from paper \citep{1981ApJ...251..604C}, the radial velocity curve of a compact object has been determined using the \ion{He}{2} 4686\,{\AA} emission line. The semiamplitude of this radial velocity curve was found to be $\sim195$ km\,s$^{-1}$ corresponding to the mass function of the compact component: 
 \begin{equation}
 f_{\rm X}(m)=\frac{m_{\rm V}^3\sin^3 i}{(m_{\rm X}+m_{\rm V})^2}\approx10.1{\rm M_\odot}.\nonumber
 \end{equation}
Here, $m_{\rm V}$ and $m_{\rm X}$ are the masses of the optical star and the compact object, respectively, and $i\approx 79^\circ$ is the binary inclination angle estimated from the moving emission lines.
Observations carried out by \cite{1990A&A...240L...5F} confirmed the result of \cite{1981ApJ...251..604C} and reported the mass function of SS433 $f_{\rm X}(m)\approx8{\rm M_\odot}$ based on the \ion{He}{2} 4686\,{\AA} emission. On the other hand, from measurements of the \ion{He}{2} 4686\,{\AA} emission line, \cite{1991Natur.353..329D} obtained a significantly lower semiamplitude of the radial velocity curve $K_{\rm X}\simeq 112$ km s$^{-1},$ corresponding to $f_{\rm X}(m)\approx2{\rm M_\odot}.$

Conclusions about the nature of the compact object in SS433 made by different authors based on high-quality spectroscopic observations are contradictory.  For example, \cite{1991Natur.353..329D} claimed that the relativistic object in SS433 is a neutron star.  The results of \cite{2010ApJ...709.1374K} also do not exclude a neutron star in SS433. According to  \cite{2004ApJ...615..422H} and  \cite{2020A&A...640A..96P}, the compact object in SS433 can be a relatively low-mass black hole $\sim 4 {\rm M_\odot}$.
Recently, in the SS433 system, a secular increase in the orbital period was discovered at a rate of $(1.14\pm0.25)\times10^{-7}$ s\,s$^{-1}$ \citep{Cher21}. The orbital angular momentum balance taking into account wind outflow from the supercritical accretion disk in SS433 implies that the mass ratio of the components should be $q=m_{\rm X}/m_{\rm V}>0.8,$ and the mass of the compact object is $m_{\rm X}>8{\rm M_\odot}$. Moreover, the presence of a neutron star with mass $<2.5 {\rm M_\odot}$ in SS433 is rejected because, in that case, the orbital period of the system should decrease, which contradicts observations \citep{2023NewA..10302060C}. Throughout the paper, we use the newest ephemeris \citep{2023NewA..10302060C}, which in $2020-2025$ can be approximated with a constant period. Thus, the orbital phases of our observations are calculated as
\begin{equation}
\varphi_{\rm orb}=({\rm HJD} - 2458933.136)/13.08341.
\end{equation}
Historical data should be phased with actual values for the initial epoch and period.
The precessional phases are calculated following \citet{2011PZ.....31....5G}:
\begin{equation}
\varphi_{\rm prec}=({\rm HJD} - 2450003)/162.284.
\end{equation}

The ambiguity in the inference about the mass of the compact object in SS433 from spectroscopic data may be due to the fact that a relatively short series of spectral observations was used. In this paper, we study the behavior of the \ion{He}{2} line by carrying out an one order of magnitude longer series of homogeneous spectroscopic observations of SS433 than in the previous studies. This enables us to average random deviations in the position and intensity of the line in order to construct the mean radial velocity curve from the \ion{He}{2} emission at different phases of the 162 day precessional period.

\section{Observations}

We have performed spectral observations of SS433 during 388 nights in $2020-2025$ using the 2.5 m telescope of the Caucasian Mountain Observatory of Sternberg Astronomical Institute of Moscow State University (CMO SAI MSU) with a low-resolution Transient Double-beam  Spectrograph (TDS) \citep{Potanin20}. The observations were carried out with a slit of 1$^{\prime\prime},$ which provides a spectral resolving power near the \ion{He}{2} line of about $\lambda/{\rm FWHM}\approx 1000$ and an accuracy of radial velocity measurements of $\sim20$ km s$^{-1}$. Each observation consisted of three frames with 300 s exposure, which resulted in a signal-to-noise ratio of about 30 at 4500\,{\AA}. In addition to our spectral data, we have used photometric BVRcIc observations of SS433 performed on the 60 cm RC600 telescope of CMO SAI MSU \citep{Berdnikov2020}.

\section{Results}
  
\begin{figure}
   \centering
   \includegraphics[width=\hsize]{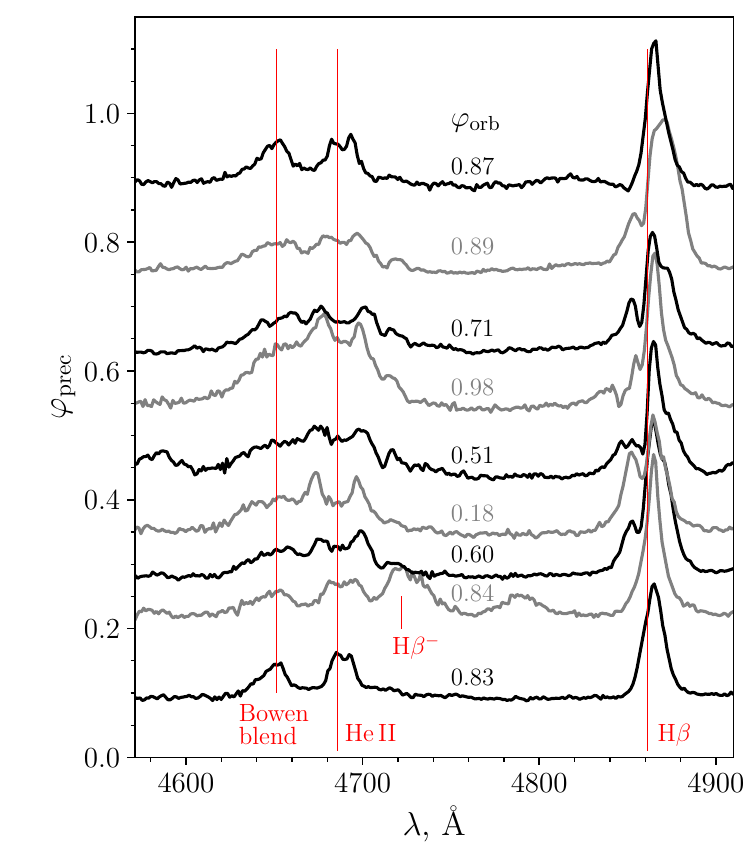}
      \caption{Examples of TDS spectra of SS433 at dates with double-peaked  \ion{He}{2} 4686\,{\AA} line.}
         \label{fig:double}
   \end{figure}

\begin{figure}
\centering
\includegraphics[width=\hsize]{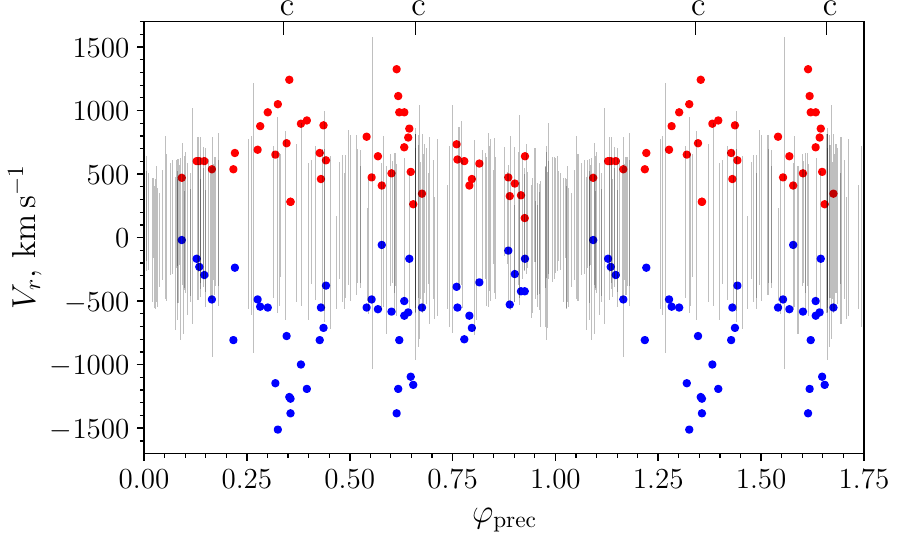} 
      \caption{The appearance of double-peaked profiles as a function of the precessional phase. Red and blue dots mark the velocities of each peak. Gray lines mark the position and FWHM of all single-peaked profiles used for constructing the radial velocity curve shown in Figure~\ref{fig:orbcurve}. Crossover phases (disk edge-on) are marked with ``c'' labels at the top. }
      \label{fig:twopeak}        
\end{figure}

\begin{figure}
   \centering
   \includegraphics[width=\hsize]{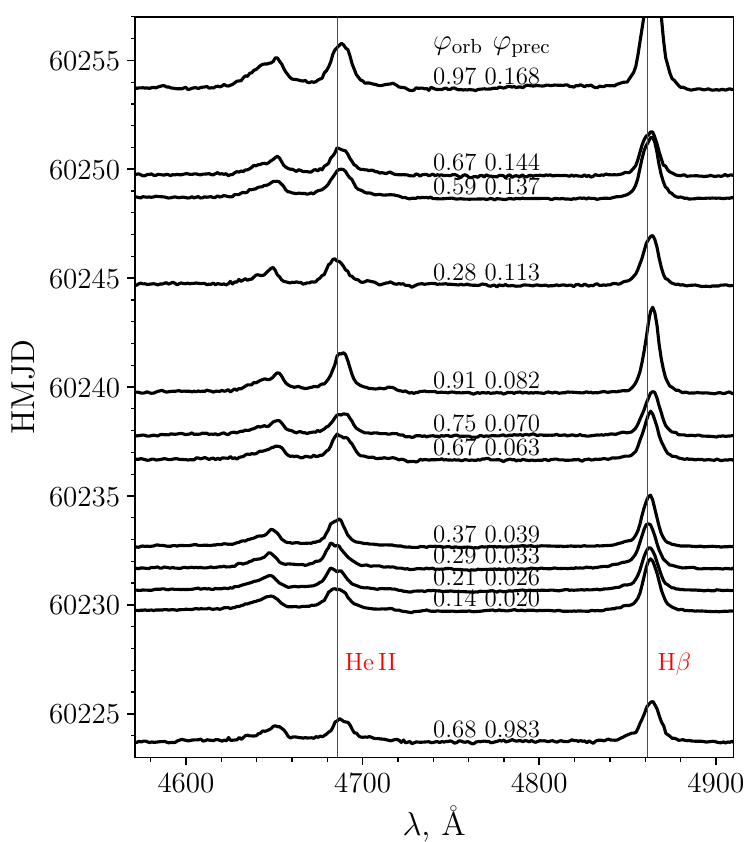}
      \caption{Examples of TDS spectra of SS433 used for measurements of radial velocities of \ion{He}{2}.}
         \label{fig:spectra}
   \end{figure}

Higher-resolution spectrographs than TDS revealed the double-peak structure of the \ion{He}{2} 4686\,{\AA} line profile \citep{1991Natur.353..329D, 2020A&A...640A..96P}. With the lower TDS resolution, these peaks usually merge, but at certain moments the line profile widens and splits with a separation between the peaks of up to 2500 km s$^{-1}$ (see Figure~\ref{fig:double} and Figure~\ref{fig:twopeak}). A similar broadening is observed on the same dates in the hydrogen and neutral helium lines.
For radial velocity measurements, we selected only those dates where no moving line is superimposed on the stationary \ion{He}{2} emission and where the profile's middle can be uniquely determined by fitting a single Gaussian. Most of these cases occur at the precessional phases near the maximum disk opening  to the viewer (moment T3, $\varphi_{\rm prec}=0$); however, profiles with regular shape can occur in other phases, too. Examples of the profiles used for the radial velocity measurements are shown in Figure~\ref{fig:spectra}.

\begin{figure}
\centering
\includegraphics[width=\hsize]{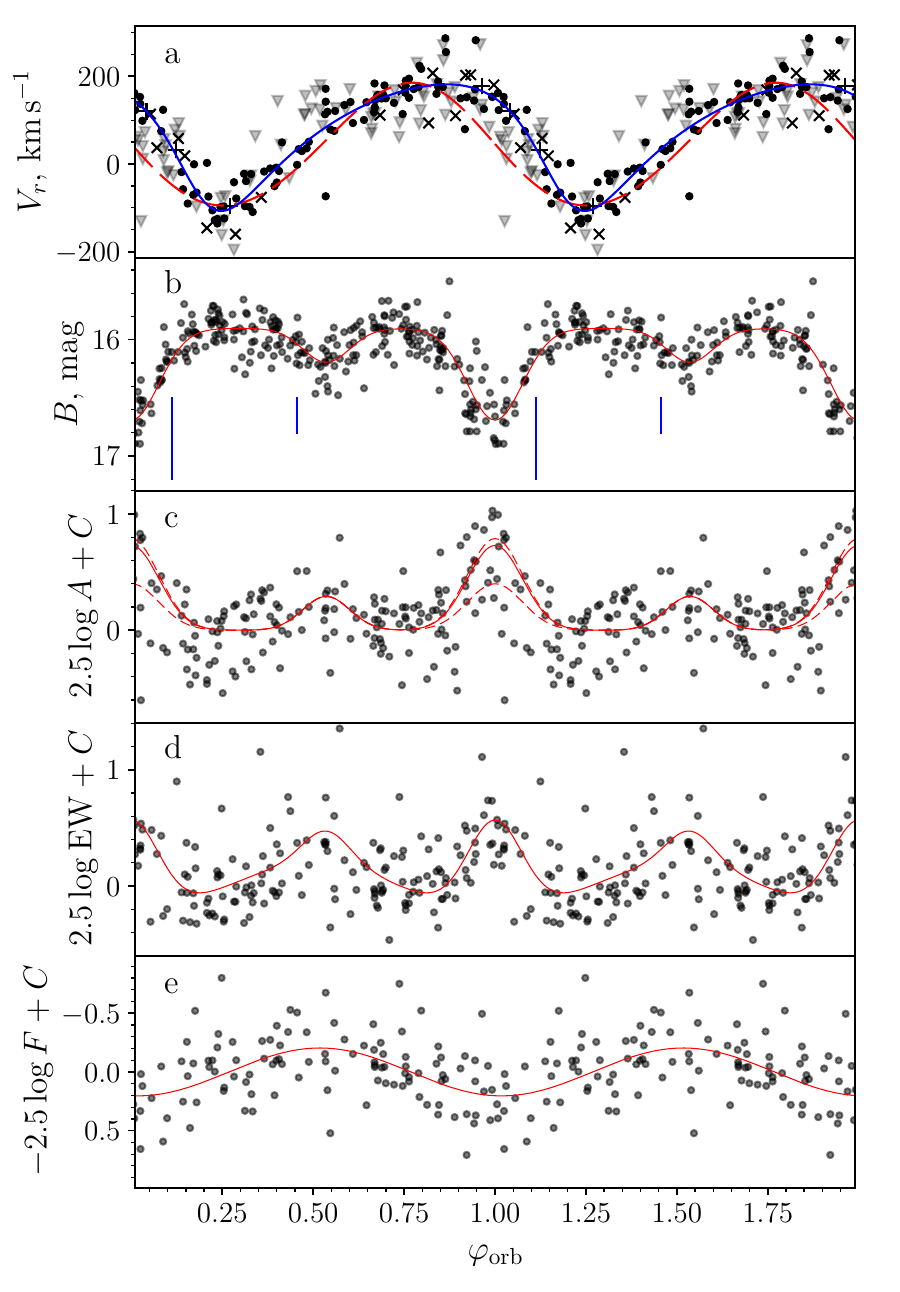} 
      \caption{Orbital phase curves of \ion{He}{2} 4686\,{\AA} line variability. (a) The radial velocity curve: black circles show our TDS observations at phases $-0.2<\varphi_{\rm prec}<0.2;$ gray triangles show our TDS observations at phases $0.2<\varphi_{\rm prec}<0.8;$ crosses ``$\times$'' are our measurements from the X-shooter spectra, plus signs ``+'' are measurements from \cite{2010ApJ...709.1374K}, the blue line is a formal fit to the radial velocities 
      ($e=0.30,$ 
      $K_{\rm X}=144$ km\,s$^{-1}$, 
      $\omega=142^\circ,$ 
      $\gamma=71$ km\,s$^{-1}$);
      the red dashed curve corresponds to the radial velocities for the orbital parameters from \cite{Cher21}
      ($e=0.05,$ $\omega=40^\circ$) scaled and shifted with $K_{\rm X}=140$ km\,s$^{-1}$ and  $\gamma=40$ km\,s$^{-1}$;
      (b) The photometric B light curve at phases $-0.2<\varphi_{\rm prec}<0.2$, the red curve is the Gaussian fits of the primary and secondary minima. Blue vertical lines indicate the would-be positions of the primary and secondary minima for the formal fit to the \ion{He}{2} radial velocity curve shown in blue in panel (a). (c) The \ion{He}{2} 4686\,{\AA} line height relative to the continuum expressed in stellar magnitudes for ease of comparison with the light curve. The red solid curve is the ``upside-down'' red light curve shown in panel (b), scaled to best match with the observed change in the line height, which corresponds to the 6\% eclipsed fraction of the \ion{He}{2} line region. The upper dashed curve shows the expected change in the line height in the absence of an eclipse. The lower dashed curve corresponds to the 30\% eclipsed fraction of the \ion{He}{2} line flux. (d) The same as in panel (c)  for the line equivalent width; the solid curve corresponds to no line eclipse, but takes into account a sinelike flux variability shown in panel (e). (e) Orbital variability of the \ion{He}{2} line flux with a sinelike fit (red curve). }
      \label{fig:orbcurve}        
\end{figure}

\begin{figure}
\centering
\includegraphics[width=\hsize]{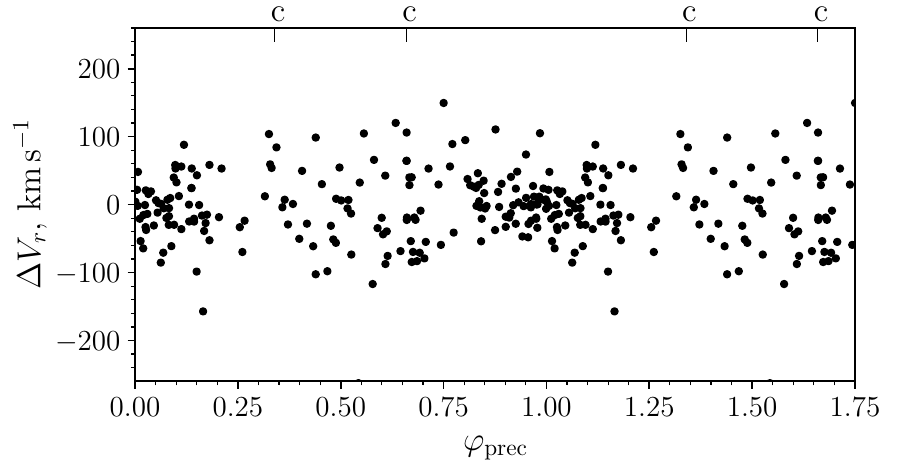} 
      \caption{Deviations of the radial velocity of \ion{He}{2} line from the formal fit in Figure~\ref{fig:orbcurve}a as a function of the precessional phase. No regular phase dependence is seen. Crossover phases are marked with ``c'' labels at the top.}
      \label{fig:diff}        
\end{figure}

In Figure~\ref{fig:orbcurve}a, we plot 92 measurements at phases $ -0.2<\varphi_{\rm prec}<0.2$ and 66 at the remaining phases. The radial velocity curve demonstrates a random spread apparently related to the multicomponent and variable structure of the line profile that is not resolved by the TDS spectrograph. The magnitude of these variations depends on the precessional phase and is minimal at phases around $\varphi_{\rm prec}=0$ and $\varphi_{\rm prec}=0.5$. The average radial velocity curve does not exhibit statistically significant changes at different precessional phases, which were noted in previous works (see, for example, \citealt{1990A&A...240L...5F}). Only a slight increase in the probability of appearance of wide double-peaked profiles can be suspected at the crossover precessional phases when the disk is visible edge-on (see Figure~\ref{fig:twopeak}). However, we stress that there is no regular dependence of deviations of the \ion{He}{2} radial velocity from the mean value on the precessional phase (see Figure~\ref{fig:diff}). 

\section{Discussion}

It can be assumed that the discrepancies in the parameters of the radial velocity curve SS433 obtained by various authors are associated with a small number of spectra used in the presence of natural velocity fluctuations due to distortions of the \ion{He}{2} line profile. For comparison, we have superimposed our radial velocity curve with measurements from \cite{2010ApJ...709.1374K}, as well as our with own measurements from archive X-shooter\footnote{\url{https://archive.eso.org/scienceportal }}  spectra of SS433 \citep{2011A&A...536A.105V}, which are described in \cite{2020A&A...640A..96P}. These data are consistent with our measurements within a random dispersion.

\subsection{Peculiarity of the radial velocity curve from \ion{He}{2} emission}

The number of points in the radial velocity curve obtained is already sufficient to see that it does not have a sinelike form, as was assumed in all previous studies. The differences from the sinelike form might be related to the SS433 orbital eccentricity $e=0.05$, which we measured in \cite{Cher21} using the eclipsing light curve. However, by taking the orbital parameters from \cite{Cher21} we get an obvious difference between the observations and the expected radial velocity curve for any amplitude $K_{\rm X}$ and $\gamma$-velocity (see red dashed curve in Figure~\ref{fig:orbcurve}a for $K_{\rm X} = 140$ km\,s$^{-1}$ and $\gamma=40$ km\,s$^{-1}$). The formal solution of the observed radial velocity curve shown by the solid line in Figure~\ref{fig:orbcurve}a yields a much higher eccentricity $e=0.30\pm0.03$ ($K_{\rm X}=144\pm4$ km\,s$^{-1}$, $\omega=142^\circ\pm6^\circ,$ $\gamma=71\pm3$ km\,s$^{-1}$, the periastron phase at $\varphi_{\rm orb}=0.191\pm0.015,$ and $f_{\rm X}(m)\approx 3.6 \rm M_{\odot}$). However, the primary and secondary minima in the corresponding light curve for this radial velocity curve would occur at $\varphi_{\rm orb}=0.11\pm0.02$ and $\varphi_{\rm orb}=0.46\pm0.03$ instead of the observed $\varphi_{\rm orb}=0.0000\pm0.0008$ and $\varphi_{\rm orb}=0.5231\pm0.0012$ (see the blue marks in Figure~\ref{fig:orbcurve}b). Thus, if we interpret the obtained radial velocity curve as a consequence of the orbital motion of the line generation region, there would be a clear contradiction with the position of the  minima on the light curve. As in the case of hydrogen lines (see the Introduction), this suggests that the radial velocity curve constructed from the \ion{He}{2} line measurements does not reflect the true orbital motion of the compact object in SS433. For orbital parameters derived from the light curve of SS433, the discrepancy between the observed radial velocity curve from \ion{He}{2} line measurements and the calculated radial velocity  curve (red dashed line in Figure~\ref{fig:orbcurve}a) can be explained by additional factors, such as relative intensity change of the multicomponent line profile or additional emission (e.g., produced in matter flowing out from the external Lagrangian point and forming the circumbinary disk; see \cite{2023MNRAS.519.1409L}) also modulated by the orbital period. But such deviations would make the line unacceptable for the measurement of the compact object's orbital motion.

\subsection{Insignificant eclipse of the \ion{He}{2} line formation region}

It is usually assumed that the formation region of the \ion{He}{2} 4686\,{\AA} line is associated with the accretion disk and/or the base of the relativistic jet. In this case, the line flux should exhibit eclipses at the phase $\varphi_{\rm orb}=0$ because with the orbital inclination $i=79^\circ$ in SS433, the limb of the optical star should screen the accretion disk center at the mid-eclipse. \citet{2010ApJ...709.1374K} noted the profile variability near $\varphi_{\rm orb}=0,$ which is attributed to the partial eclipse of the line formation region. However, our observations do not show systematic changes in the profile and/or the Rossiter-McLaughlin effect at phases $\varphi_{\rm orb}\approx0.$  

To check whether the line formation region is eclipsed,  we examined the change in the line height above the continuum in Figure~\ref{fig:orbcurve}c, as well as the equivalent width (EW) of the line in Figure~\ref{fig:orbcurve}d (the latter showed a greater dispersion due to the inaccuracy of the line width $\sigma$ estimates when fitting it with a single Gaussian; however, it has a more clear physical meaning, so we present both curves). The dispersion of points on the plots is primarily related to the actual irregular change in the line flux, see, for example, Figure~3 from \cite{2020A&A...640A..96P}, where the line flux varies by a factor of two at the same precessional and orbital phases. From Figures \ref{fig:orbcurve}cd it can be seen that the line's height and EW increase during both the primary and secondary eclipse minima in strict anticorrelation with the light curve. This is a reliable indication that the line formation region is not eclipsed substantially. The near-constant emission line flux appears stronger relative to the weakening continuum during eclipses, that is to say, such variability of the line height and EW reflects the contrast of the (quasi)constant emission line on top of the variable continuum.

For a quantitative description, we fitted the primary minimum on the light curve with a Gaussian, and then found the height of the Gaussian with the same position and width for the \ion{He}{2} line height change. From the ratio of amplitudes of the Gaussians, we found that the optimal value of the eclipsed part of the line is $6\%\pm7\%$ (solid line in Figure~\ref{fig:orbcurve}c) that is, the data can be consistent even with the complete absence of the eclipsed part (upper dashed line in Figure~\ref{fig:orbcurve}c). The largest fraction of the eclipsed line flux cannot exceed 30\% (lower dashed line in Figure~\ref{fig:orbcurve}c). Thus, most of the \ion{He}{2} 4686\,{\AA} line flux is produced far from the accretion disk. The line  profile and average radial velocity curve constructed from its measurements carry information not about the orbital motion of the disk, but about the velocity field in an extended gas shell that is periodically perturbed by the binary system immersed in it. The multicomponent structure of the profile, resolved at higher resolution, suggests that the variability of $V_r$ can arise not only as a result of the orbital Doppler shift of the components but also due to changes in their relative intensities leading to the line profile asymmetry and formal shift of the effective line center. However, most flux from these line components is not eclipsed.

\subsection{Orbital variability of the absolute \ion{He}{2} emission flux}

Due to light slit losses, the observed line fluxes can be underestimated by a factor weakly dependent on the wavelength. We can estimate this factor from broadband photometry by comparing the observed fluxes in some filter with those calculated from the spectra by integrating them with the filter transmission curve and detector sensitivity curve. More than half of our spectra are accompanied by photometric measurements taken within $\pm5$ hr that allow us to renormalize them. Based on these data, we constructed the dependence of the absolute line flux on the orbital phase and found that the flux varies almost sinusoidally with a semiamplitude of 0.2 mag: the line turns out to be the brightest at the phase $\varphi_{\rm orb}=0.5,$ when the disk is in front of the donor star (Figure~\ref{fig:orbcurve}e). This variability cannot be caused by screening of a substantial part of the line forming region inside the binary system, because in this case the flux is expected to drop somewhere within the phases $\varphi_{\rm orb}=0.75-0.25$ and a plateau outside this range should appear, which is inconsistent with the observed sinelike flux variability in Figure~\ref{fig:orbcurve}e. Taking into account this variability and assuming that there are no eclipses of the \ion{He}{2} 4686\,{\AA} line formation region, we superimposed the expected change in the equivalent width on its  measurements  and also obtained excellent agreement (cf. Figure~\ref{fig:orbcurve}d).

Note that the sinelike change in the \ion{He}{2} 4686\,{\AA} line flux can be traced at different precessional phases, but at $\varphi_{\rm prec}=0$ the average flux is by $\approx10$\% higher than at the crossover phases and by $\approx 15$\% higher than at $\varphi_{\rm prec}=0.5.$ A similar variability, but with a larger amplitude of $\sim100$\%, was observed in the  H$\alpha$ emission line \citep{2022ARep...66..451C}.

\section{Conclusion}
\label{sec:conclusion}
Using the results of 6 yr of spectral and photometric monitoring of SS433 in $2020-2025$ at CMO SAI MSU, we have found that the formation region of the \ion{He}{2} 4686\,{\AA} emission line can experience only weak eclipses suggesting that the line formation region is significantly larger than the donor star size. The absolute flux in the line against the randomly fluctuating background demonstrates a regular sinelike variability with the orbital period of SS433 and maximum around phase 0.5 (Figure~\ref{fig:orbcurve}e). The double-peaked profiles of the \ion{He}{2} emission line and the nature of their variability are similar to those in the hydrogen and neutral helium emission lines, which were described by \citet{2011A&A...531A.107B} in the model of a rotating circumbinary envelope with azimuthal inhomogeneity. Such an inhomogeneity should naturally occur in the form of a spiral shock wave in the circumbinary gas due to the orbital motion of the components.

Our findings imply that it is incorrect to consider the radial velocity of the \ion{He}{2} 4686\,{\AA} line as the radial velocity of the compact object, and therefore, the mass function estimates from the previous high-quality \ion{He}{2} 4686\,{\AA} line measurements may not correspond to real values. The observed \ion{He}{2} 4686\,{\AA} radial velocity curve (Figure~\ref{fig:orbcurve}a) cannot be used to reliably estimate the masses of the binary components in SS433.

\begin{acknowledgements}
The authors acknowledge the anonymous referee for useful comments. We thank the staff of the CMO SAI MSU for their help with observations. The study was conducted under the state assignment of M.~V.~Lomonosov Moscow State University. Scientific equipment used in this study was purchased partially through the M.~V.~Lomonosov Moscow State University Program of Development.
\end{acknowledgements}

\section*{Data availability}
All observational data (spectra, B-photometry, and parameters of the \ion{He}{2} line shown in Figures \ref{fig:twopeak} and \ref{fig:orbcurve}) are available at \dataset[doi:10.5281/zenodo.18311703]{\doi{10.5281/zenodo.18311703}}.
 
\begin{contribution}
All authors contributed equally.
\end{contribution}

%
\facilities{SAI-2.5m, VLT:Kueyen}

\software{astropy \citep{2013A&A...558A..33A,2018AJ....156..123A,2022ApJ...935..167A},  
          }



\bibliography{ss433he}{}
\bibliographystyle{aasjournalv7}



\end{document}